\documentclass[12pt]{article}
\usepackage{graphicx}
\usepackage{natbib}

\newcommand\pubnumber{Article 36 in eConf C1304143}
\newcommand\pubdate{\today}

\def\asu{$^{{\rm  1}}$School of Earth and Space Exploration
Arizona State University, Tempe, AZ 85281, USA \\
$^{{\rm  2}}$ Department  of  Physics  and  Astronomy,
  University  of  Leicester, University Road,  Leicester, LE1 7RH, UK \\
$^{{\rm  3}}$ Department of  Physics,   University  of   Warwick,  Coventry,
  CV4 7AL, UK \\}

\def\Title#1{\begin{center} {\Large #1 } \end{center}}
\def\Author#1{\begin{center}{ \sc #1} \end{center}}
\def\Address#1{\begin{center}{ \it #1} \end{center}}

\newcommand\pubblock{\rightline{\begin{tabular}{l} \pubnumber\\
         \pubdate  \end{tabular}}}
\newenvironment{Abstract}{\begin{quotation}  }{\end{quotation}}
\newenvironment{Presented}{\begin{quotation} \begin{center}
             PRESENTED AT\end{center}\bigskip
      \begin{center}\begin{large}}{\end{large}\end{center} \end{quotation}}
\def\Acknowledgements{\bigskip  \bigskip \begin{center} \begin{large}
             \bf ACKNOWLEDGEMENTS \end{large}\end{center}}
\def\beq{\begin{equation}}
\def\eeq#1{\label{#1}\end{equation}}
\def\eeqn{\end{equation}}

\def\beqa{\begin{eqnarray}}
\def\eeqa#1{\label{#1}\end{eqnarray}}
\def\eeqan{\end{eqnarray}}

\let\bar=\overbar

\def\Dslash{\not{\hbox{\kern-4pt $D$}}}
\def\dslash{\not{\hbox{\kern-2pt $\del$}}}

\def\msb{{\bar{\ssstyle M \kern -1pt S}}}

\begin{document}
\begin{titlepage}
\pubblock

\vfill
\Title{Are GRBs the  same at  high and  low  redshift?}
\vfill
\Author{Owen Littlejohns$^{{\rm   1,2}}$, Nial Tanvir$^{{\rm  2}}$, Richard Willingale$^{{\rm  2}}$, Paul O'Brien$^{{\rm  2}}$, Phil Evans$^{{\rm  2}}$, Andrew Levan$^{{\rm  3}}$}
\Address{\asu}
\vfill
\begin{Abstract}
  Due  to their highly  luminous nature,  gamma-ray bursts  (GRBs) are
  useful tools in studying the early  Universe (up to $z$ $=$ 10).  We
  consider  whether  the   available  subset  of  \textit{Swift}  high
  redshift GRBs are unusual  when compared to analogous simulations of
  a bright  low redshift  sample.  By simulating  data from  the Burst
  Alert  Telescope   (BAT;  \citealt{2005SSRv..120..143B})  the  light
  curves of these  bright bursts are obtained over  an extensive range
  of redshifts,  revealing complicated evolution in  properties of the
  prompt emission such as $T_{90}$.
\end{Abstract}
\vfill
\begin{Presented}
GRB 2013 \\
the Seventh Huntsville Gamma-Ray Burst Symposium \\
Nashville, Tennessee, 14--18 April 2013
\end{Presented}
\vfill
\end{titlepage}
\def\thefootnote{\fnsymbol{footnote}}
\setcounter{footnote}{0}

\section{Introduction}

\begin{figure}[h]
  \centering
  \includegraphics[height=8cm]{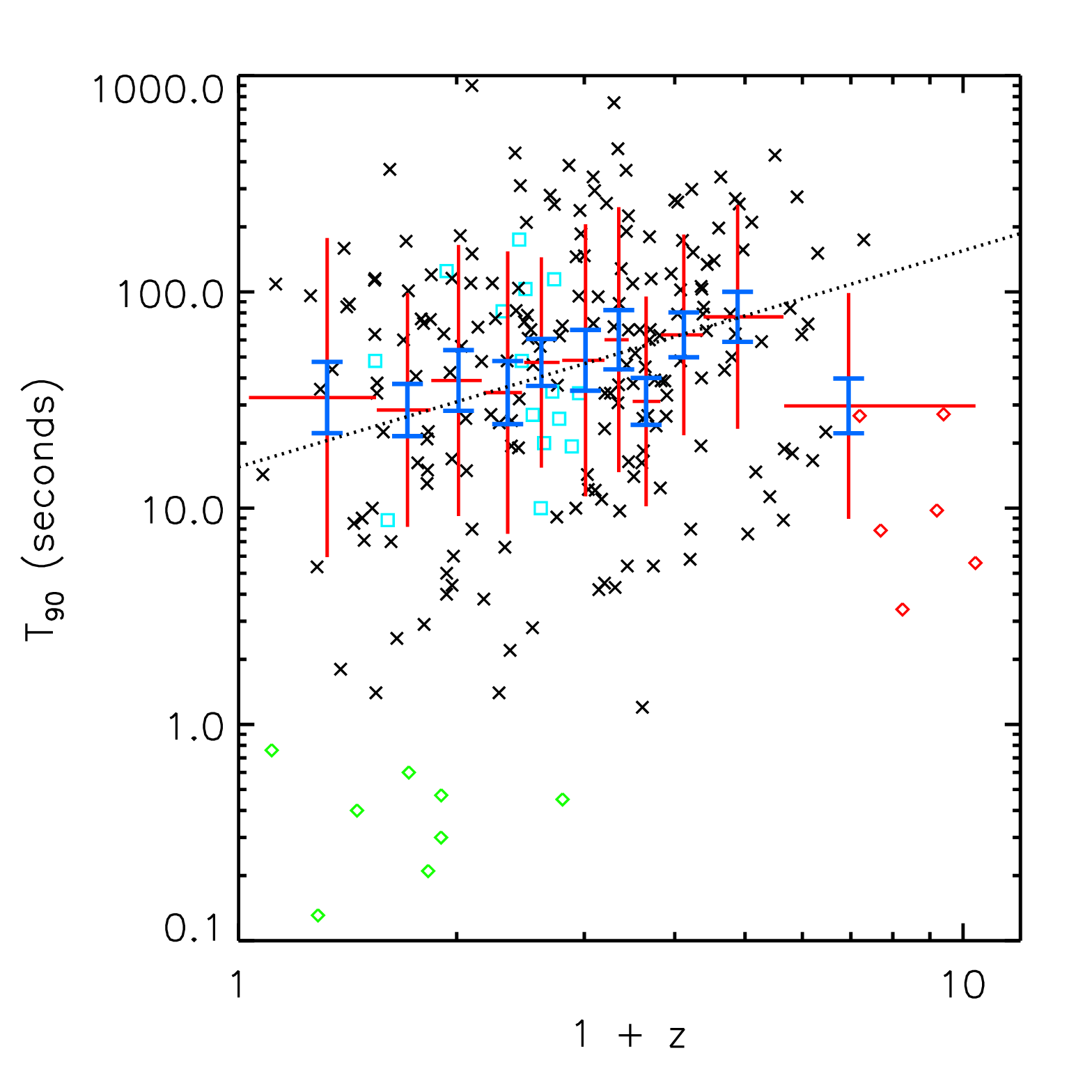}
  \caption{The   observed  distribution  of   $T_{90}$  for   the  203
    \textit{Swift} \citep{2004ApJ...611.1005G}  bursts with a redshift
    estimate prior to the 15$^{{\rm th}}$ of July 2012.  Also included
    are GRB 120521C and GRB 120923A which are both candidates for high
    redshift.  The red crosses show geometric averages of bursts taken
    with 20 bursts  in each bin (except in the final  bin which has 16
    bursts). The red  error bars shown in the  vertical direction show
    the     root    mean     square    (RMS)     scatter    calculated
    logarithmically. Also  shown are the  standard errors on  the mean
    for each  bin in  blue. The dotted  black line shows  the expected
    evolution  due  to   simple  cosmological  time  dilation,  namely
    $T_{90}$  $\propto$ $1+z$.  Short  GRBs are  denoted by  the green
    diamonds. The high-redshift subset of 6 GRBs shown in Figure 2 are
    indicated by red diamonds.}
  \label{fig:t90nosht}
\end{figure}

\begin{figure}[h]
  \centering
  \includegraphics[height=8cm]{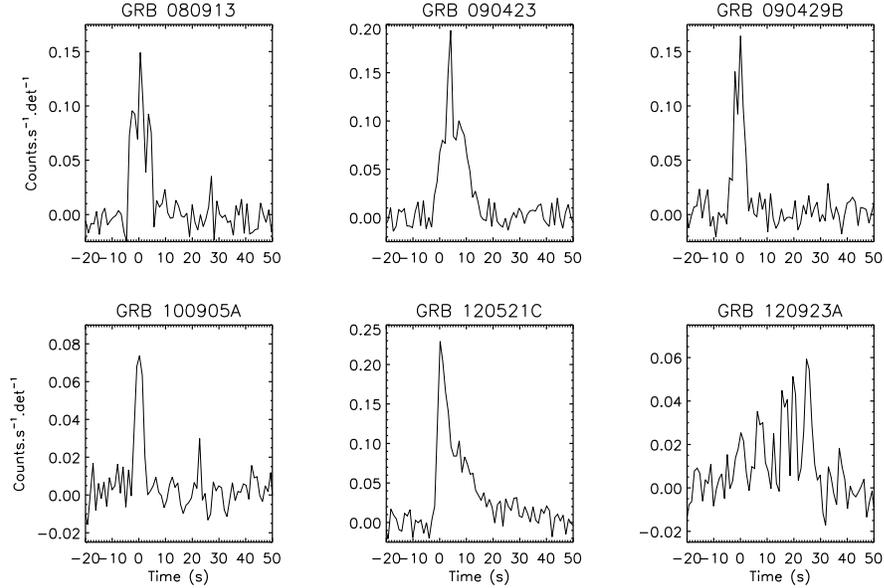}
  \caption{15--350 keV light curves  of rate-triggered GRBs within the
    high-redshift subset.  The light curves  are binned at 1024  ms to
    allow clearer identification of the structure within each.}
  \label{fig:lchiz}
\end{figure}

An  important  question  in  the   field  of  GRBs  is  whether  their
populations  change  with  redshift.   This, in  principle,  might  be
reflected in the typical prompt behaviour. Beyond the dearth of higher
redshift short ($T_{90}$ $<$ 2  s) bursts, the only tentative evidence
for an  evolution in  the population of  long-GRBs is that  a majority
(6/7)  of the highest  redshift ($z$  $>$ 6)  GRBs found  to-date have
apparently rather  short durations, $T_{90}/\left(1+z\right)$  $<$ 5 s
(see Figures \ref{fig:t90nosht}  \& \ref{fig:lchiz}).  Such values are
of  particular interest  as they  approach  the classical  value of  2
seconds used as  a crude estimate of whether a burst  is long or short
(\citealt{1993ApJ...413L.101K},        although        see        also
\citealt{2013ApJ...764..179B}).\par

\section{Method}

The creation of the simulated light curves required accounting for the
effects of cosmological time dilation; of band shifting, as individual
photons  are  redshifted  (this  included  consideration  of  the  BAT
response and  change in  background as a  function of energy);  and of
declining flux due to increased luminosity distance.\par

We  used  model  fits   obtained  using  the  methodology  with  which
\citet{2010MNRAS.403.1296W}    applied     the    pulse    model    of
\citet{2009MNRAS.399.1328G}.   The characteristic times,  energies and
normalising  fluxes for each  pulse were  calculated at  the simulated
redshift using Equations \ref{eq:first} to \ref{eq:third}.\par

\begin{equation}
  T_{{\rm sim}} = \left( \frac{1+z_{{\rm sim}}}{1+z_{{\rm orig}}} \right)
  T_{{\rm orig}},
  \label{eq:first}
\end{equation}
\begin{equation}
  E_{{\rm pk,sim}} = \left( \frac{1+z_{{\rm orig}}}{1+z_{{\rm sim}}} \right)
  E_{{\rm pk,orig}},
  \label{eq:second}
\end{equation}
\begin{equation}
  S_{{\rm pk,sim}} = \left( \frac{K_{{\rm orig}}}{K_{{\rm sim}}} \right)
  \left( \frac{D_{{\rm L,orig}}}{D_{{\rm L,sim}}} \right)^{2} S_{{\rm pk,orig}},
  \label{eq:third}
\end{equation}

where  $T$ denotes a  characteristic time,  $z$ is  redshift, $E_{{\rm
    pk}}$ is the peak energy of each pulse spectrum, $S_{{\rm pk}}$ is
the normalising flux, $K$ is the k-correction and $D_{{\rm L}}$ is the
luminosity distance.\par  Noise was added to the  light curves, before
using a  rate-triggering algorithm  designed to establish  whether BAT
would trigger  on the  observed flux. Those  bursts which  were bright
enough to  create a  trigger were then  run through the  standard {\sc
  battblocks} software to obtain $T_{90}$.\par

\section{Results}

\begin{figure}
  \centering
  \includegraphics[height=8cm,angle=270]{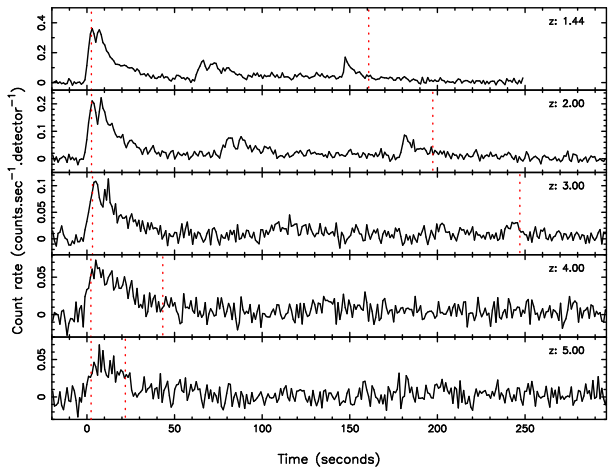}
  \caption{Simulated 15--350 keV light curves of GRB 100814A ($z_{{\rm
        orig}}$ $=$  1.44) at a  variety of redshifts. The  binning of
    these light curves  is 1024 ms, and the red  dotted line shows the
    start and  end of the  identified $T_{90}$ period.  Note  that the
    vertical scale of each panel  is independent to allow structure to
    be seen in all panels.}
  \label{fig:100814A_fig}
\end{figure}

Example simulated  light curves  for GRB 100814A  are shown  in Figure
\ref{fig:100814A_fig}.  The  top  three  panels show  an  increase  in
measured  $T_{90}$  with  redshift.   At  higher  redshifts  late-time
structure is no longer detectable, thus $T_{90}$ reduces.\par

Unlike  previous work  \citep{2013ApJ...765..116K} the  simulated GRBs
contained multiple pulses.  As such three effects were apparent in the
$T_{90} \left( z  \right)$ evolution: the time dilation  of pulses and
intervening  quiescent  periods,  the  gradual loss  of  the  emission
``tail'' of  individual pulses  and the total  loss of  fainter (often
late-time) pulses.\par

\begin{figure}
  \centering
  \includegraphics[height=8cm]{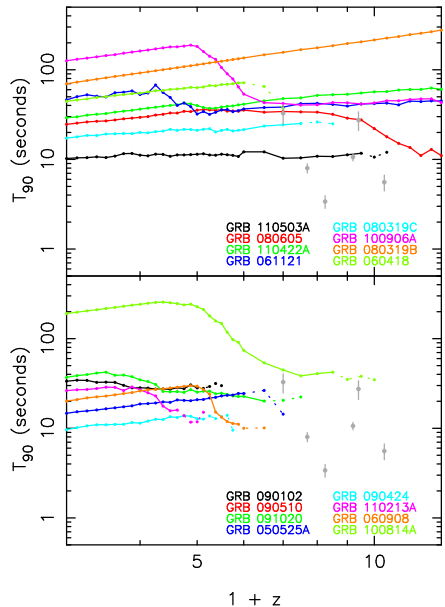}
  \caption{The  evolution of $T_{90}$  (averaged over  100 simulations
    per burst) as  a function of redshift.  The  sample has been split
    according to the luminosity of  the brightest pulse in each burst,
    with  the most luminous  8 being  shown in  the top  panel.  Solid
    lines show where  the bursts was detected in a  minimum of 90\% of
    the repeated  simulations, the dotted  lines show where  the burst
    was  detected in less  than 90\%  but greater  than 50\%  of these
    repeats.   Each burst is  represented by  a different  colour data
    set.   The grey  points  correspond to  the  15--350 keV  $T_{90}$
    values of the observed high redshift bursts.}
  \label{fig:t90evo}
\end{figure}

Figure \ref{fig:t90evo} shows the impact of these competing effects on
duration  and also  provides a  good comparison  between  the observed
high-redshift  sample  and  simulations  of  the  bright  low-redshift
bursts.  Even taking  the 16  bursts  with the  brightest features  at
low-redshifts, only approximately half of these remain detected at $z$
$>$ 6.  This confirms  that the detected  high-redshift bursts
only represent the luminous tail of the total population.\par

The observed  high-redshift sample contains  four bursts in the  5 $<$
$T_{90}$ $<$ 10  s regime and now also two with  $T_{90}$ $\sim$ 30 s.
For those simulated bursts which  are still visible at $z$ $>$
6, the typical duration is 30 $<$ $T_{90}$ $<$ 70 s. This suggests the
two populations are  different, although the analysis is  limited by a
small sample size.\par

\section{Conclusions}

Predicting the evolution of  $T_{90}$ with redshift is complicated. It
is  a  combination of  cosmological  time  dilation  of structure  and
intervening quiescence,  gradual loss of  the pulse ``tails''  and the
total  loss of fainter  structure.\par

Simulating bright low-redshift bursts at high redshifts show that only
the very brightest  structure is detected.  While the  two most recent
high-redshift  candidates  are  longer  in duration,  the  simulations
typically have longer $T_{90}$ values.\par

Using the  simulation technique developed  with more GRBs  detected at
high-redshift will allow us to  fully answer whether GRBs are the same
at high-redshift.\par

\Acknowledgements

This work is supported at the University of Leicester by the STFC.

%\begin{thebibliography}{99}
%\bibitem
%\end{thebibliography}

\begin{bibliography}{t90Nashville}
  \bibliographystyle{mn2e}
\end{bibliography}

%%
%%  bibliographic items can be constructed using the LaTeX format in SPIRES:
%%    see    http://www.slac.stanford.edu/spires/hep/latex.html
%%  SPIRES will also supply the CITATION line information; please include it.
%%
\end{document}